\documentclass[%
 amsmath,amssymb,superscriptaddress
]{revtex4-1}

\usepackage{graphicx}
\usepackage{dcolumn}
\usepackage{bm}
\usepackage[mathlines]{lineno}

\begin{document}

\title{High accuracy determination of the $^{238}$U/$^{235}$U fission cross section ratio up to $\sim$1 GeV at n\_TOF (CERN) }

\author{C.~Paradela} \affiliation{Universidade de Santiago de Compostela, Spain}\affiliation{European Commission JRC, Institute for Reference Materials and Measurements, Retieseweg 111, B-2440 Geel, Belgium}
\author{M.~Calviani} \affiliation{CERN, Geneva, Switzerland}
\author{D.~Tarr{\'{\i}}o} \affiliation{Universidade de Santiago de Compostela, Spain}\affiliation{Department of Physics and Astronomy, University of Uppsala, Sweden}
\author{E.~Leal-Cidoncha} \affiliation{Universidade de Santiago de Compostela, Spain}
\author{L.S.~Leong} \affiliation{Centre National de la Recherche Scientifique/IN2P3 - IPN, Orsay, France}\affiliation{Japan Atomic Energy Agency, JAEA, Japan}
\author{L.~Tassan-Got} \affiliation{Centre National de la Recherche Scientifique/IN2P3 - IPN, Orsay, France}
\author{C.~Le Naour} \affiliation{Centre National de la Recherche Scientifique/IN2P3 - IPN, Orsay, France}
\author{I.~Duran} \affiliation{Universidade de Santiago de Compostela, Spain}
\author{N.~Colonna} \email[]{nicola.colonna@ba.infn.it}
\author{L.~Audouin} \affiliation{Centre National de la Recherche Scientifique/IN2P3 - IPN, Orsay, France}
% ...

\collaboration{The n\_TOF Collaboration (www.cern.ch/ntof)}  \noaffiliation

\date{\today}

\begin{abstract}

The $^{238}$U to $^{235}$U fission cross section ratio has been determined at n\_TOF up to $\sim$1 GeV, with two different detection systems, in different geometrical configurations. A total of four datasets have been collected and compared. They are all consistent to each other within the relative systematic uncertainty of 3-4\%. The data collected at n\_TOF have been suitably combined to yield a unique fission cross section ratio as a function of the neutron energy. The result confirms current evaluations up to 200 MeV.  A good agreement is also observed with theoretical calculations based on the INCL++/Gemini++ combination up to the highest measured energy. The n\_TOF results may help solving a long-standing discrepancy between the two most important experimental dataset available so far above 20 MeV, while extending the neutron energy range for the first time up to $\sim$1 GeV.

\begin{description}
\item[PACS numbers]
\pacs{}25

\keywords{}

\end{description}

\end{abstract}

\maketitle


\section{Introduction}

The neutron-induced fission cross section of the two major isotopes of Uranium, $^{235}$U and $^{238}$U, are of fundamental importance in the field of nuclear technology, as well as for other fields of fundamental and applied Nuclear Physics. In particular, fission cross section data above a few MeV are important for the development of new systems for energy production and waste transmutation, for accelerator and space applications, in measurements of high-energy neutron flux, or for the refinement of theoretical models on nuclear fission at high energy. The $^{235}$U(\textit{n,f}) cross section is a standard at 0.0253 eV and from 0.15 eV to 200 MeV \cite{carlsondatasheet,nds5}, while the $^{238}$U(\textit{n,f}) is recommended for use as a standard in the neutron energy region between 2 and 200 MeV \cite{nds8}.  While the $^{235}$U(\textit{n,f}) cross section is commonly used in a variety of fields, for example in neutron flux measurements, from thermal to very high energy, the $^{238}$U(\textit{n,f}) can be more conveniently used in the presence of a low energy neutron background, thanks to its high fission threshold.

Despite its importance, few data have been collected up to now on the $^{238}$U(\textit{n,f})/$^{235}$U(\textit{n,f})
cross section ratio above 20 MeV. Lisowski \textit{et al.}~\cite{liso} measured the ratio at Los Alamos National Laboratory, from 0.5 to 400 MeV. Current evaluated nuclear data libraries~\cite{endf} and evaluations of cross section standards~\cite{carlsondatasheet} are mostly based on these results. More recently, Shcherbakov \textit{et al.}~\cite{shcherba}, measured the ratio between 1 and 200 MeV at the GNEIS facility, Gatchina. Above 50 MeV these results are between 5 and 8\% lower than those of Lisowski \textit{et al.}. On the contrary, results from Nolte \textit{et al.}~\cite{nolte} seem to confirm the older measurement, but they are affected by a too large systematic uncertainty to draw a final conclusion. New, high accuracy measurements are therefore required to solve the long-standing discrepancy, and improve the accuracy in current evaluations. An extension of the energy range up to higher energy would also be desirable, as data above 200 Me could prove useful for refining theoretical models, in particular those used in modern Monte Carlo codes for high-energy neutron transport. 

To address the need of new, accurate data to improve current cross section standard, and to extend the current energy limit, a series of measurements of the $^{238}$U(\textit{n,f})/$^{235}$U(\textit{n,f}) cross section ratio were performed at the n\_TOF facility at CERN, from 0.5 MeV to 1 GeV, in different experimental campaigns and with different setups. After a description of the various setups and analysis procedures, in Section II, the n\_TOF results are presented and compared with each other in Section III. Suitably combined in a unique dataset, the n\_TOF results are then compared with previous experimental data, with current evaluations and with the predictions of a theoretical model extending up to the highest measured energy. Conclusions are finally given in Section IV.

\section{Experimental setup}

The measurements were performed at the n\_TOF facility at CERN~\cite{ntof,perfrep,guerre,barba}, in different experimental campaigns and with different experimental setups, taking advantage of the very convenient features of the neutron beam, that make this facility particularly suitable for measurements of neutron-induced fission cross sections~\cite{me}. In particular, the wide energy distribution of the neutron beam, extending over more than ten orders of magnitude, allows one to measure the fission cross section up to the GeV region. Furthermore, the long flight path, close to 200 m, ensured a good energy resolution at all energies. 

The $^{238}$U(\textit{n,f})/$^{235}$U(\textit{n,f}) cross section ratio can be determined from the background-subtracted fission events C(E$_{n}$), as a function of neutron energy, according to the following expressions:
\begin{equation}
\frac{\sigma_{8}(E_{n})}{\sigma_{5}(E_{n})} = \frac{C_{8}(E_{n})}{C_{5}(E_{n})}\times\frac{N_{5}}{N_{8}}\times\frac{\epsilon_{5}}{\epsilon_{8}}\times CF
\label{eq:one}
\end{equation}

Here the subscripts \textit{8} and \textit{5} refer to the $^{238}$U(\textit{n,f}) and $^{235}$U(\textit{n,f}) reactions, respectively, \textit{N} is the areal density of the samples used in the measurements (in atoms/barn), $\epsilon$ the detection efficiency for fission events, and \textit{CF} accounts for other correction factors, in particular for the detection dead-time.
It is important to note that the ratio does not depend neither on the shape nor on the absolute value of the neutron flux, since the two samples, having the same diameter, are measured simultaneously and in the same experimental setup, i.e. are exposed to the same neutron flux. As a consequence, the final uncertainties in the cross section ratio will mostly be determined by the uncertainty on the areal density of the two samples, and on the ratio of the efficiency and dead-time corrections.

The data reported here were collected with two detection systems: a Fission Ionization Chamber (FIC), in which a single Fission Fragment (FF) is detected, and an array of Parallel Plate Avalanche Counters (PPAC), in which the two FF are detected in coincidence. This last detection system was used in two geometrical configurations, i.e. perpendicular to the neutron beam direction, and in a tilted configuration, with the detectors mounted at an angle of 45$^{\circ}$ relative to the neutron beam direction. A brief description of the detectors and other experimental details is given below for each setup, together with the systematic uncertainties affecting the corresponding results. All data here reported were collected with a large-aperture collimator (8 cm diameter). With such an arrangement, the beam profile is essentially flat over the surface of samples with diameter $\leq$8 cm. This feature is of great advantage in this as in other high-accuracy fission cross section measurements at n\_TOF, since it minimizes the effect of possible inhomogeneities of the fissile deposit.
Another important aspect of the n\_TOF neutron beam regards the neutron energy resolution in the range reported here. Above a few MeV, this is dominated by the width of the proton beam, of 6 ns (r.m.s.). Considering the long flight path of 185 meters, the energy resolution is less than 1\% at 100 MeV, reaching slightly over 5\% at 1 GeV.

\subsection{The FIC chamber}

The fast ionization chamber used in the measurement is described in detail in Ref.~\cite{marconim}. It is made of a stack of ionization chambers mounted along the direction of the neutron beam. Each cell consists of a central aluminum electrode 100 $\mu$m thick, plated on both sides with the isotope to be measured, and two external 15-$\mu$m-thick aluminum electrodes. Different versions of the FIC were built. One of them, the FIC1, was specifically built for measurements of highly radioactive samples, and according to CERN regulation had to comply with the ISO2919 standard, as "sealed source". For this reason, the case was made in stainless steel, and the chamber was equipped with thick stainless steel windows. The second version of the ionization chamber, FIC2, was much lighter and was directly coupled to the vacuum beam pipe, with only a 125 $\mu$m kapton window at the gas/vacuum interface. For this reason, this second version was less sensitive to the so-called $\gamma$-flash, i.e. the prompt signal caused in the detector by spallation $\gamma$-rays and relativistic particles. In both FIC1 and FIC2 chambers the gap between electrodes was 5 mm. All chambers were operated with a gas mixture of 90\% Ar and 10\% CF$_{4}$ at a pressure of 720 mbar. The detector signals were amplified by a current feedback operational amplifier AD844 and digitized with a flash analog-to-digital converter (ADC) with 8-bit resolution and 250 MHz sampling rate. The fast timing properties of the gas and of the front-end electronics resulted in a fast signal with a typical rise time of 40 ns and a fall time of 120 ns. 

\begin{table}[t!]
\caption{List of samples used in the measurement with the FIC1 and FIC2 fission ionization chambers. The samples had a diameter of 8 and 5 cm for the FIC1 and FIC2, respectively.}
\label{tab1}
\begin{tabular}{ccccc}
\hline\hline
\textbf{Setup}&\textbf{Sample}&\textbf{Mass}&\textbf{Areal density}&\textbf{Uncertainty}\\
 & &(mg)&(atoms/b)&(\%)\\
\hline
& $^{235}$U & 15.2 & 7.75$\times$10$^{-7}$ & 1.4\\
& $^{235}$U & 16.6 & 8.46$\times$10$^{-7}$ & 1.3\\
& $^{238}$U & 12.8 & 6.44$\times$10$^{-7}$ & 1.4\\
FIC1& $^{238}$U & 12.4 & 6.24$\times$10$^{-7}$ & 1.4\\
& $^{238}$U & 13.4 & 6.74$\times$10$^{-7}$ & 1.2\\
& $^{238}$U & 13.7 & 6.90$\times$10$^{-7}$ & 1.4\\
\hline
& $^{235}$U & 6.47 & 8.44$\times$10$^{-7}$ & 1.1\\
FIC2& $^{235}$U & 6.32 & 8.25$\times$10$^{-7}$ & 1.1\\
& $^{238}$U & 20.0 & 2.6$\times$10$^{-6}$ & $>$3\%\\
\hline\hline
\end{tabular}
\end{table}

Two different measurements of the $^{238}$U/$^{235}$U fission cross section ratio were performed with FIC1 and FIC2, respectively. In both measurements, the chambers were positioned at approximately 190 m downstream of the spallation target. In the measurement with FIC1, two $^{235}$U and four $^{238}$U samples were mounted, in the form of U$_{3}$O$_{8}$ compound. The samples were prepared at the Institute of Physics and Power Engineering, Obninsk, Russia by means of the painting technique. Their purity, checked by means of $\alpha$-spectroscopy, was in all cases greater than 99\%, with only trace contamination of other U isotopes. The uniformity of the fissile deposit was always between 5 and 10\%.The deposit of fissile material was 8 cm in diameter, while the areal density of each sample was between 6$\times$10$^{-7}$ and 8$\times$10$^{-7}$ atoms/barn, with a declared uncertainty on the mass of each sample of $\sim$ 1.3\% (see Table~\ref{tab1}). Due to a large $\gamma$-flash signal, it was not possible to collect useful data with this setup above 10 MeV. These results, on the other hand, are affected by a small systematic uncertainty on the mass of each sample, and possible inhomogeneities of the samples are averaged out when combining the different samples. Therefore, these data were used as basis for the normalization of the second dataset, affected by larger systematic uncertainties.

The second setup used in the measurement consisted of the FIC2 chamber, inside which two $^{235}$U and one $^{238}$U samples were mounted. All three samples had a diameter of 5 cm. The areal density of the two $^{235}$U samples was around 6$\times$10$^{-7}$ atoms/barn, while a much thicker $^{238}$U sample, of 2.6$\times$10$^{-6}$ atoms/b was used in this second setup, to collect a sufficient statistics below the fission threshold as well. The uncertainty on the mass of this last sample was $>$3\%, and no information was available on the homogeneity of the deposit. For this reason, the data collected with this setup had to be normalized to the results obtained with the FIC1 chamber, in the 1-10 MeV neutron energy region. The characteristics of all samples used in the measurement with the fission ionization chamber are listed in Table \ref{tab1}. 

For the FIC2 chamber, an electronic compensation technique was applied in order to extract the FF signals from the tail of the $\gamma$-flash at short time-of-flights, i.e. for neutron energies above approximately 10 MeV. The technique is based on the observation that the shape of the $\gamma$-flash signal is similar for contiguous electrodes. Therefore, by subtracting the output of two consecutive electrodes, one of them without sample, it is possible to extract clear fission fragment signals. The presence of a residual tail related to the $\gamma$-flash required a more careful reconstruction procedure, with the threshold on the amplitude varying as a function of the neutron energy. The maximum energy reached with this procedure was around 200 MeV.

In order to extract the neutron energy from the measured time-of-flight, a calibration was performed using the resonances in the $^{235}$U(n,f) cross section. Corrections for the neutron beam attenuation in the samples and electrodes, and for the divergence of the beam profile, were estimated to be less than 1\%, and were therefore neglected. Similarly, corrections for the sample inhomogeneities were considered negligible, in particular since the n\_TOF neutron beam, for the large aperture collimator, shows a nearly flat spatial profile. On the contrary, corrections had to be applied for the detection efficiency and dead-time effects, both of which different for the two samples. Efficiency corrections were applied only to the FIC1 data, used for normalization. The efficiency, estimated by means of detailed FLUKA simulations~\cite{fluka} of the energy deposition in the gas as a function of the threshold used in data analysis, was found to be 94.9\% and 96.2\% for the thin $^{235}$U and $^{238}$U samples mounted in FIC1, respectively. Contrary to the efficiency, dead-time corrections had to be applied to both FIC1 and FIC2 data, since they depend both on the sample and neutron energy. Such corrections, based on the non-paralyzable model, were below 5\% in the whole neutron energy range for both thin samples, while it reached $\sim$10\% for the thick $^{238}$U sample used in FIC2.


\subsection{The PPAC setup}

Fission cross sections have been measured at n\_TOF with the coincidence technique, using a detection system based on Parallel Plate Avalanche Counters. The detectors are described in detail in Refs.~\cite{carlospara, diegotarrionim}. They are characterized by very thin windows, small gaps between electrodes (3 mm) and low pressure of the gas. These features make the detector practically insensitive to the prompt $\gamma$-flash. Furthermore, the signals produced by the FF are very fast, less than 10 ns in width, so that the pile-up probability is very small. Most importantly, the very fast timing combined with the low sensitivity of the detectors to the $\gamma$-flash, allows one to record signals at very short time-of-flights, or equivalently at very high neutron energies, nominally up to 1 GeV.

\begin{table}[b!]
\caption{Characteristics of the samples used with the PPAC setup, both in the perpendicular configuration and in the first measurement with the tilted configuration ("PPAC tilted 1"). }
\label{tab2}
\begin{tabular}{ccccc}
\hline\hline
\textbf{Sample}&\textbf{Mass}&\textbf{Areal density}&\textbf{Uncertainty}\\
&(mg)&(atoms/b)&(\%)\\
\hline
 $^{235}$U & 13.97 & 6.98$\times$10$^{-7}$ & 0.7 \\
 $^{238}$U & 11.5 & 5.86$\times$10$^{-7}$ & 0.7 \\
\hline\hline
\end{tabular}
\end{table}

A stack of Parallel Plate Avalanche Counters, placed in the beam, were used at n\_TOF for measuring FF in coincidence. The actinide samples were deposited on very thin backing and positioned between two PPACs. The main advantage of the coincidence technique is the very high efficiency for rejecting the $\alpha$-particle background related to the natural radioactivity of the samples, as well as for discriminating fission against competing reactions, in particular high-energy neutron-induced reactions on the PPAC structural material, producing recoiling nuclei and charged particles inside the detectors. Another advantage of PPACs is that they can also supply information on the angular distribution of fission fragments. Each cathode consists of 2-mm-wide strips, suitably read-out to provide a one-dimensional position information. By combining the signals from the two orthogonal cathode strips, the fission fragment trajectory can be determined, allowing to reconstruct the emission angle. The main drawback of the system is represented by a limited angular acceptance, since at relatively large angles one of the fragments looses a large fraction of its energy in traversing the sample backing and the entrance windows of the detectors, either being stopped before reaching the gas volume, or falling below the detection threshold. Therefore, while the single technique is characterized by an efficiency for detecting the single fragment close to 100\% (since the fragment has only to escape from the deposit of fissile material), in the coincidence technique the efficiency for grazing trajectories drops rapidly, going to zero for emission angles larger than $\sim$60$^{\circ}$, the exact limiting angle depending on the thickness of the sample backing. 

\begin{figure}[h!]
\includegraphics[angle=0,width=0.9\linewidth,keepaspectratio]{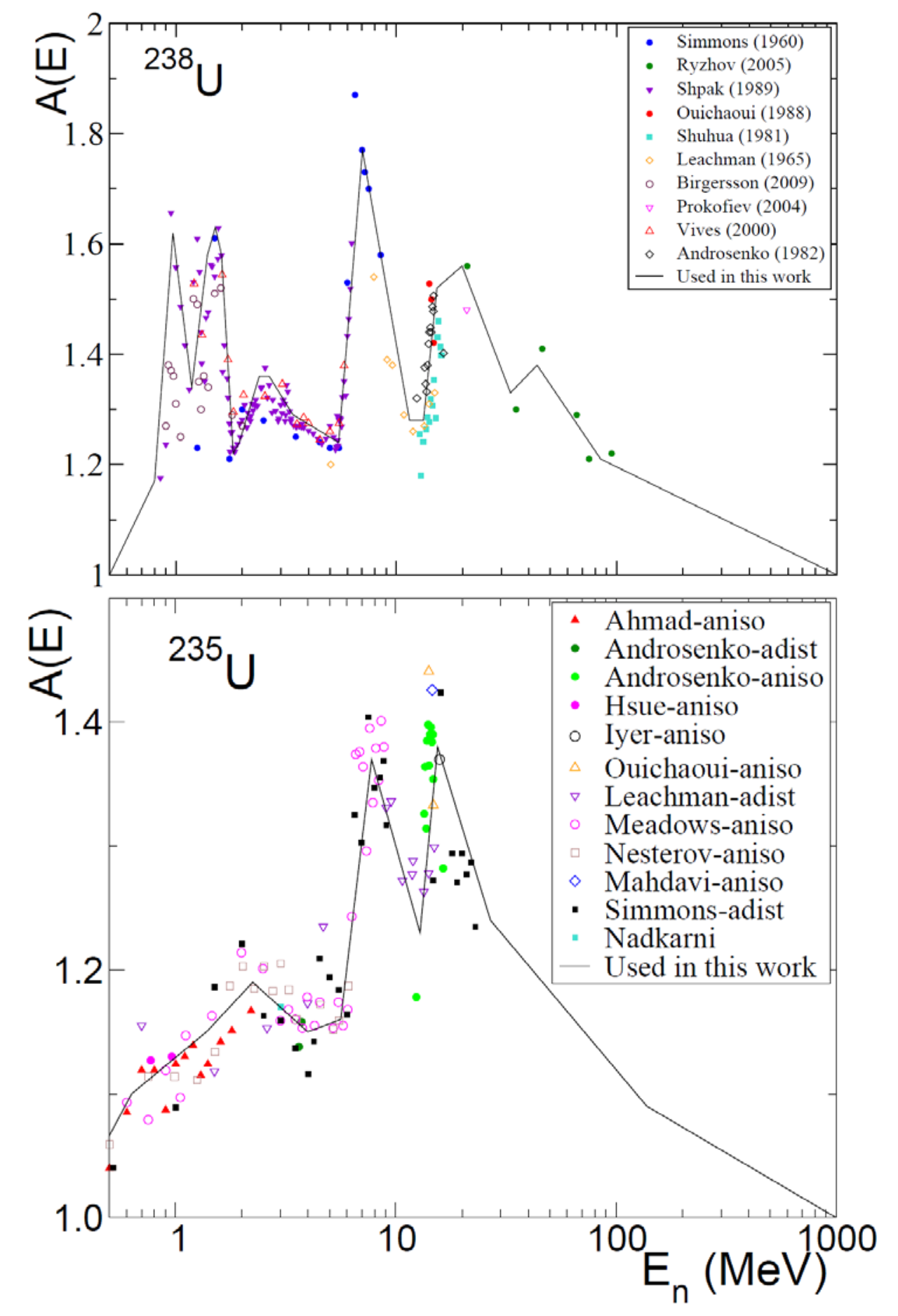}
\caption{\label{fig:anysot} (Color online) Anisotropy in the emission angle of FF in the neutron-induced fission of $^{238}$U (upper panel) and $^{238}$U (lower panel), as a function of the neutron energy. The curves represent the results of a fit of the various experimental data with Legendre polynomials.}
\end{figure}

In the first experimental campaign at n\_TOF the PPAC detectors were mounted in the neutron beam perpendicularly to the beam direction. Hereafter we will refer to this configuration as "PPAC perpendicular". As previously mentioned, in this configuration the setup is affected by a loss of efficiency for FF emitted at angles larger than $\sim$60$^{\circ}$.
To overcome the problem of the limited angular coverage, a new geometrical configuration was adopted at n\_TOF in the second experimental campaign, with the detectors and the samples mounted at 45$^{\circ}$ relative to the neutron beam direction. A scheme of the tilted configuration can be found in~\cite{diegotarrionim,diegosheet}. Two measurements were performed with the tilted setup, hereafter referred to as "PPAC tilted 1", and "PPAC tilted 2".  In all cases the samples, prepared at IN2P3-Orsay, were 8 cm in diameter, with the fissile material electro-deposited on a thin aluminum foil of 2.5 $\mu$m thickness in the first two measurements, and on a 0.7 $\mu$m Al backing in the third measurement. The characteristics of the samples used in the first two cases are listed in Table~\ref{tab2}. Their thickness and homogeneity was checked by means of the Rutherford Back Scattering spectroscopy (RBS), which also provides the chemical composition of samples and backings. Apart from a contamination of $\sim$6\% of $^{238}$U in the  $^{235}$U sample, taken into account in the analysis, only trace concentrations of other isotopes were found in the samples (see Refs.~\cite{carlospara, diegotarrioprc}). The samples of the last measurement ("PPAC tilted 2") have not been characterized with the necessary accuracy. As a consequence, this last dataset has been normalized to the ENDF/B-VII.1 evaluated ratio between 3 and 5 MeV.

\begin{figure}[h!]
\includegraphics[angle=0,width=1.\linewidth,keepaspectratio]{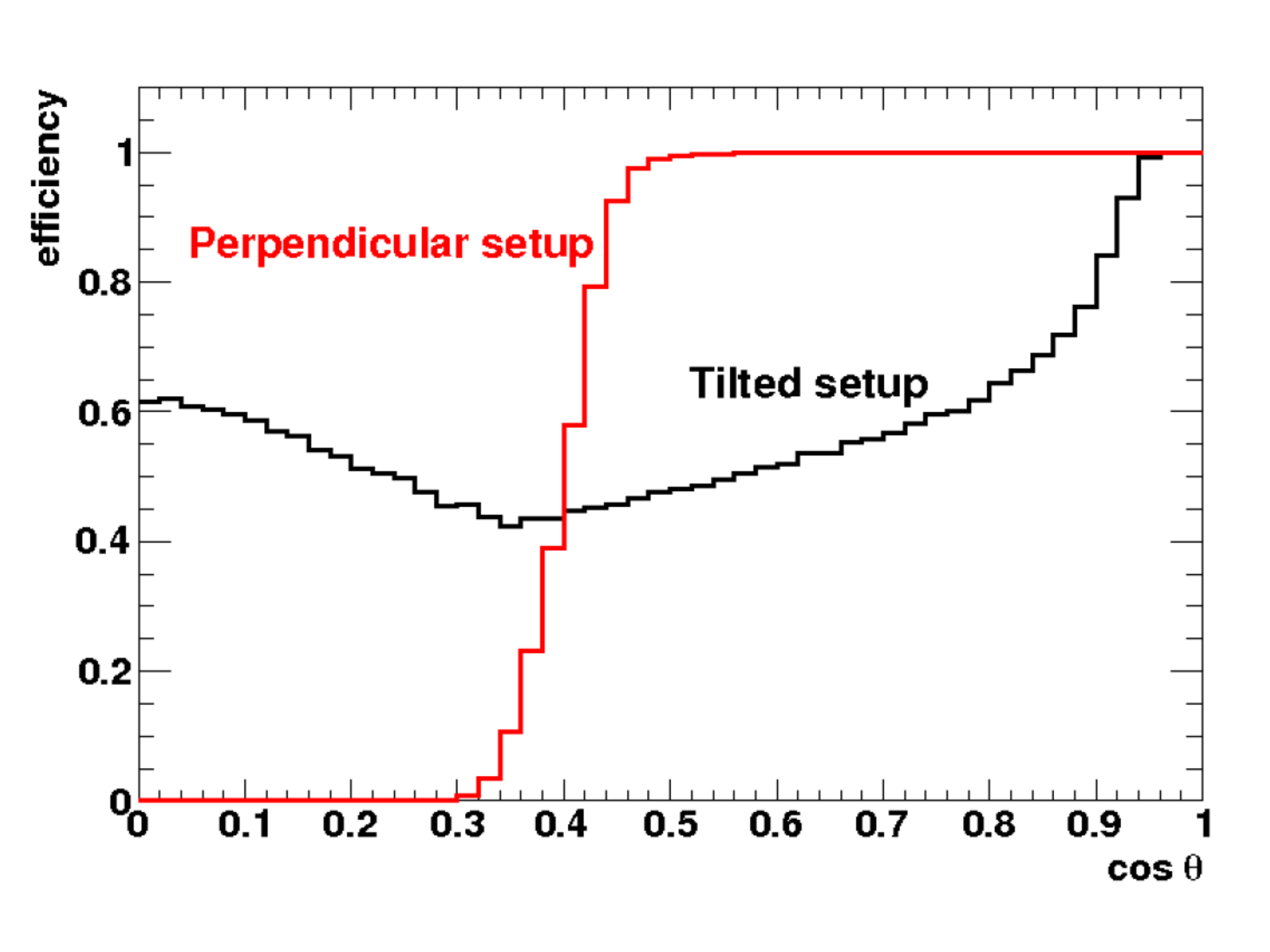}
\caption{\label{fig:efficiencyppac} (Color online) Efficiency for detecting FF in coincidence with the PPAC setup, in the two geometrical configurations. In the tilted setup, the gain of efficiency at large emission angles is compensated by a corresponding loss at all other angles.}
\end{figure}

In both perpendicular and tilted configurations, a correction had to be applied for the loss of efficiency related to the stopping of one of the fission fragments in the sample backing. Because of the angular dependence of the efficiency, the fission fragment angular distribution (FFAD) had to be considered. To this end, corrections were calculated, by means of Monte Carlo simulations, on the basis of existing experimental results on the FFAD as a function of the neutron energy, reported in the EXFOR database for both isotopes~\cite{exfor}. More details on the procedure can be found in Refs.~\cite{carlospara,diegotarrioprc}. The FFAD is expressed as the sum of even Legendre polynomials in terms of the cosine of the emission angle (because of the forward-backward symmetry of the emitted FF). In this work, polynomials up to second order only were considered, as there is no reliable information in literature about higher order coefficients. The energy-dependent coefficients were determined from the so-called anisotropy parameter, A(E), defined as the ratio between the emission probability of the FF at 0 and 90$^{\circ}$, W(E,0$^{\circ}$)/W(E,90$^{\circ}$). The angular anisotropy of $^{235}$U and $^{238}$U was obtained by fitting data present in literature, available in the EXFOR database~\cite{exfor}. Since data only go up to $\sim$20 MeV neutron energy, the anisotropy was constrained to asymptotically approach unity above this energy. The data and the results of the fit are shown in Figure~\ref{fig:anysot}.

The detection efficiency of the PPAC setup as a function of the FF emission angle is shown in Figure~\ref{fig:efficiencyppac} for the perpendicular and tilted configuration. The figure clearly shows the difference between the two configurations: in the perpendicular case, the efficiency is flat up to a given angle (around $\sim$60$^{\circ}$), and drops to zero above this limit. On the contrary,  in the tilted configuration all FF emission angles are covered. The gain in efficiency at large angles is obviously compensated by a reduction of the efficiency at smaller angles, due to the cut in the azimuthal angle related to the stopping of the fragments in the dead layers (sample backing and PPAC windows). As expected, the global efficiency, i.e. the efficiency integrated over the polar angle, is the same in both perpendicular and tilted configuration, but the latter has the advantage of a smaller influence of the angular anisotropy of FF emission. 

\begin{figure}[h!]
\includegraphics[angle=0,width=1.\linewidth,keepaspectratio]{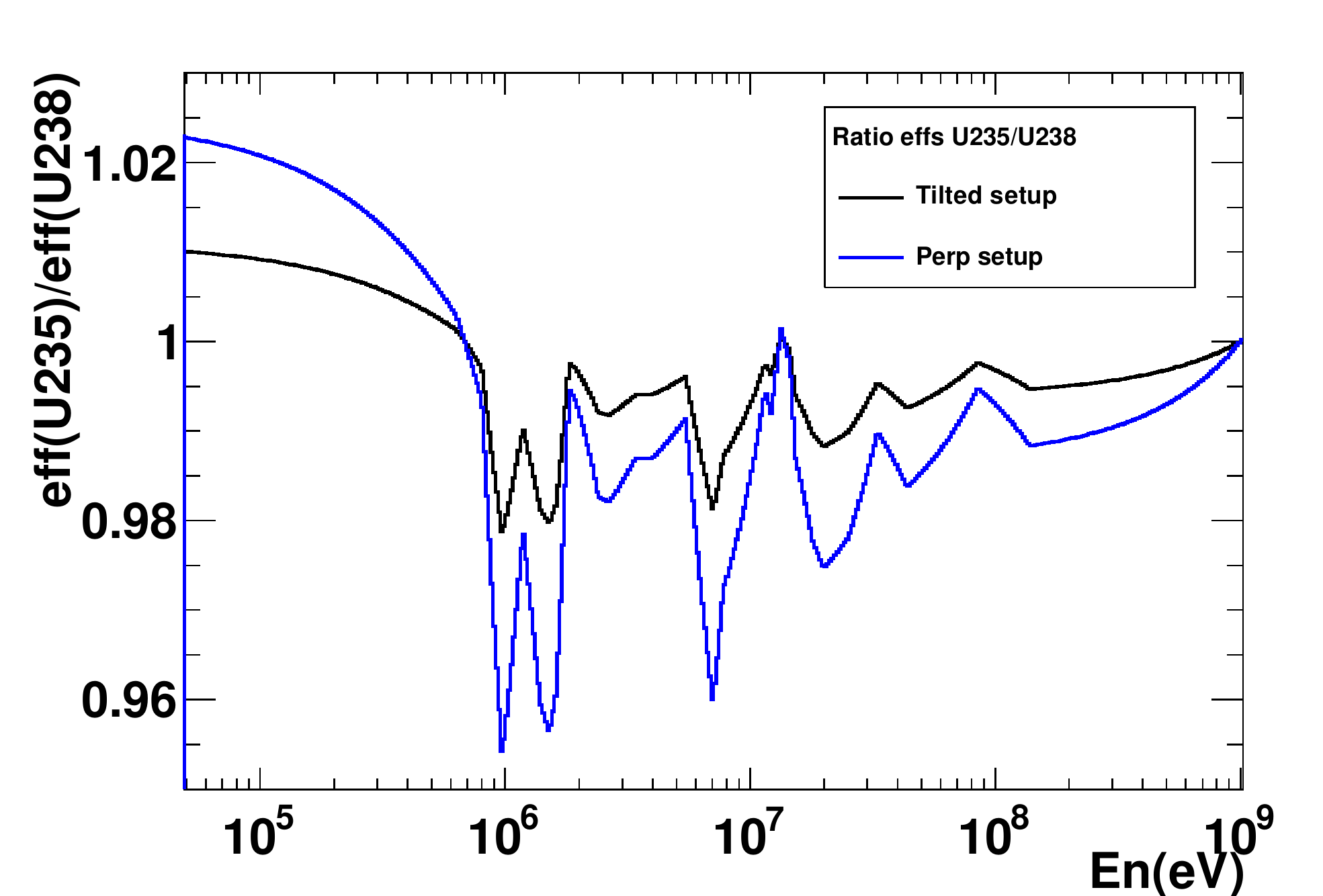}
\caption{\label{fig:ratioefficppac} (Color online) Ratio between the efficiencies for detecting a fission event from the $^{235}$U and $^{238}$U samples respectively. The various structures in the efficiency ratio corresponds to the structures in the angular anisotropy of FF emission in the two reactions.}
\end{figure}

The analysis of the tilted configuration is described in great detail in Refs.~\cite{diegotarrionim, diegotarriophd, lousaiphd}. Several conditions were applied in order to identify the fission fragments, reconstruct the position of emission from the sample and the emission angle of each fission fragment, reject all possible sources of background, etc... As previously mentioned, efficiency corrections in the tilted configuration were estimated by Monte Carlo simulations~\cite{diegotarriophd}, but in this case the results were confirmed by means of a method based exclusively on experimental data~\cite{lousaiphd}. Contrary to the perpendicular configuration, a dependence of the efficiency on the polar angle has to be taken into account, together with other effects related to the viewing angle of the sample by the detectors.

As shown in Eq.~\ref{eq:one}, the cross section ratio depends on the ratio of the efficiency for detecting a fission event from the $^{235}$U and the $^{238}$U samples respectively. The ratio is shown in Figure~\ref{fig:ratioefficppac}, for the two different configurations. In both cases, the correction is of the order of at most 4\%. The systematic uncertainties affecting the various PPAC measurements, as well as the measurement with the FIC, are listed in Table~\ref{tab3}.

\section{Results and discussion}

\begin{figure}[b!]
\includegraphics[angle=0,width=1.25\linewidth,keepaspectratio]{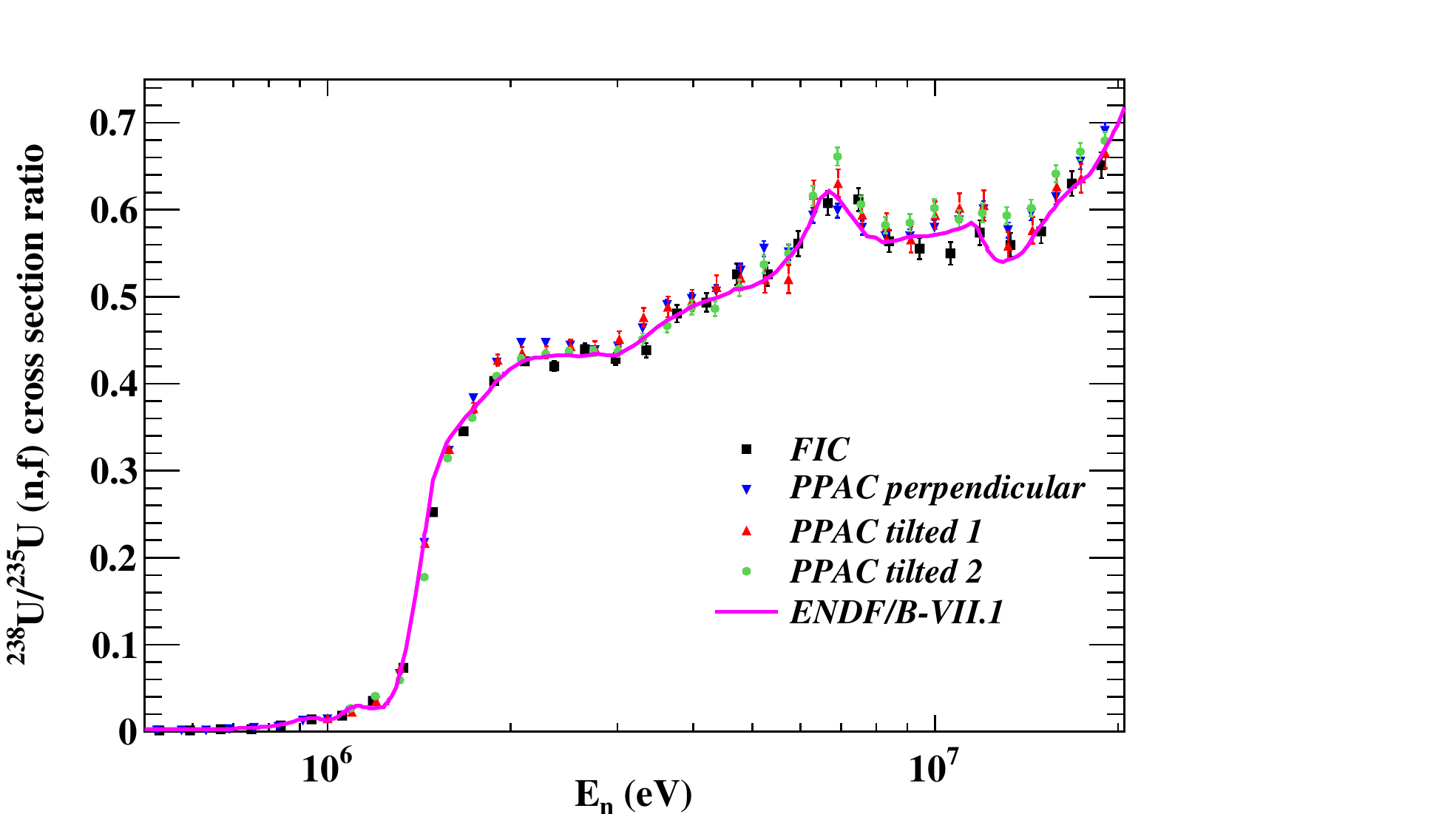}
\caption{\label{fig:ratioall20MeV} (Color online) The $^{238}$U/$^{235}$U fission cross section ratio measured at n\_TOF with the different setups, in the neutron energy range 500 keV - 20 MeV. The "PPAC tilted 2" data were normalized to the ENDF/B-VII.1 ratio between 3 and 5 MeV. The error bars represent the statistical errors only. For comparison, the ratio calculated on the basis of the evaluated cross sections from ENDF/B-VII.1 is also shown in the figure.}
\end{figure}

Figure~\ref{fig:ratioall20MeV} shows the results of the different measurements of the $^{238}$U/$^{235}$U fission cross section ratio, for neutron energies in the range 500 keV-20 MeV. For comparison, the ratio extracted from the ENDF/B-VII.1 evaluated cross sections, is also shown in the figure. We remark here that for neutron energies below 20 MeV a wealth of data are available in literature, with the standard cross section of both isotopes characterized by an uncertainty of less than 1\%. Therefore the very good agreement between n\_TOF results and evaluations, evident in Figure~\ref{fig:ratioall20MeV}, provides a strong confidence on the accuracy of the new data, in particular on the systematic uncertainty related to the sample mass and on the efficiency corrections. Some minor differences can nevertheless be noticed in the Figure, in particular around 12 MeV neutron energy, where the dip corresponding to a valley in the $^{238}$U(\textit{n,f}) cross section is less pronounced in the  n\_TOF dataset, relative to ENDF/B-VII.1 evaluation. Another minor difference can be observed around the $^{238}$U fission threshold, slightly above 1 MeV. In this region, the n\_TOF data are systematically shifted on the right, with respect to evaluations, possibly indicating a slight underestimate of the value of the fission threshold in the evaluations.

\begin{table}[t!]
\caption{Systematic uncertainties on the data collected in the four different measurements of the $^{238}$U/$^{235}$U fission cross section ratio. All values are expressed in percentage. "Others" refers mostly to dead-time corrections. For the last dataset, normalized to ENDF/B-VII.1, the uncertainty in the mass is replaced by the one in the evaluated cross sections.}
\label{tab3}
\begin{tabular}{ccccc}
\hline\hline
\textbf{Setup}&\textbf{Samples}&\textbf{Efficiency}&\textbf{Others}&\textbf{Total}\\
\hline
FIC & 2 & 1 & 3 & \textbf{3.5} \\
PPAC perpend. & 1.1 & 3 & $<$1 & \textbf{$\gtrsim$3} \\
PPAC tilted 1 & 1.1 & 2 & $<$1 & \textbf{$\lesssim$3} \\
PPAC tilted 2& ($\sim$1) & 1 & $<$1 & \textbf{2.5} \\
\hline\hline
\end{tabular}
\end{table}

\begin{figure}[h!]
\includegraphics[angle=0,width=1.25\linewidth,keepaspectratio]{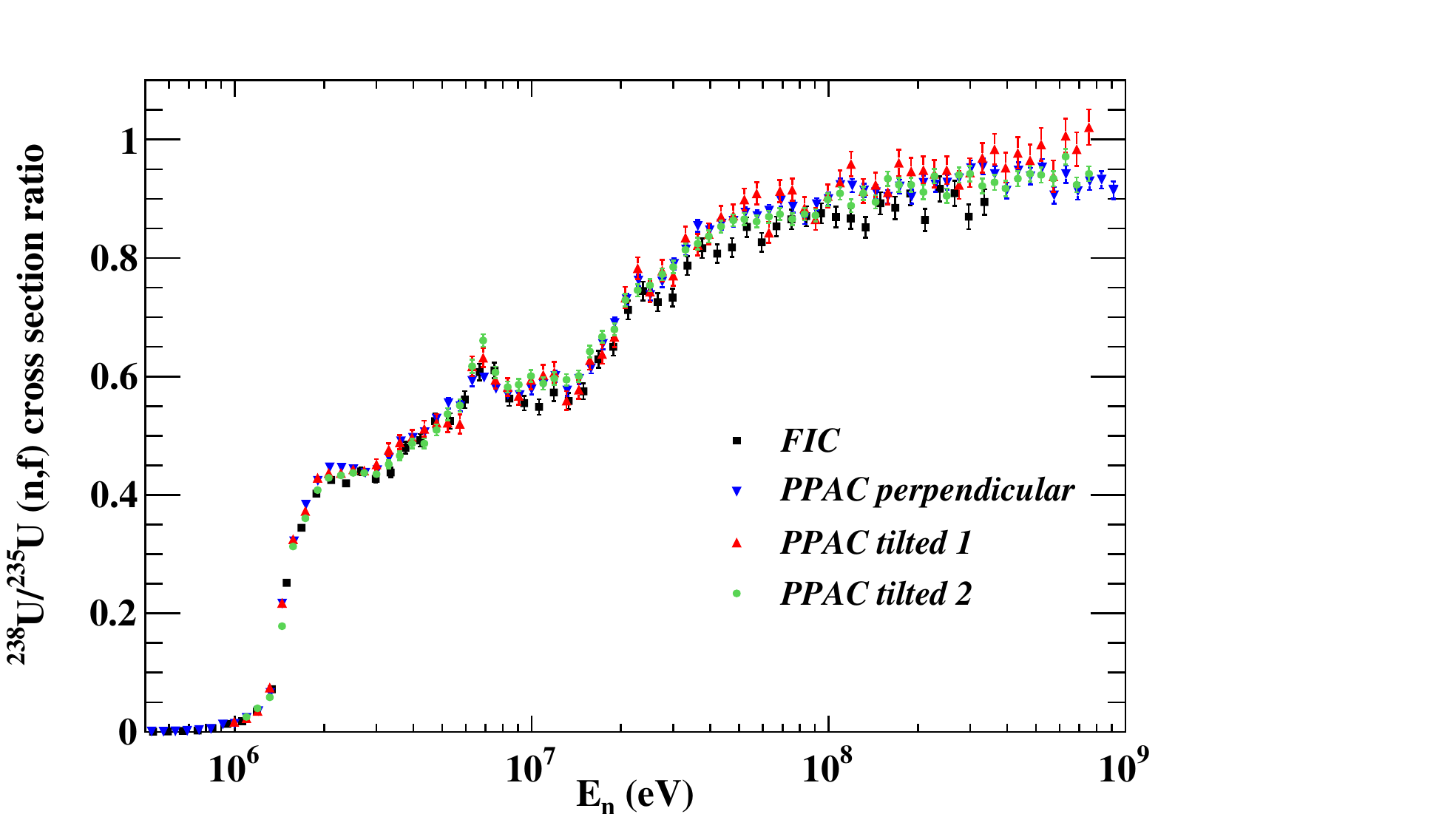}
\caption{\label{fig:ratioall} (Color online) The $^{238}$U/$^{235}$U fission cross section ratio from the different  n\_TOF measurements, in the neutron energy range 500 keV - 1 GeV.}
\end{figure}

Figure~\ref{fig:ratioall} shows the results of the four different measurements at n\_TOF, from 500 keV to 1 GeV. The error bars on the symbols indicate the statistical errors only. Above 20 MeV some differences can be observed between the various results. In particular, the FIC results are systematically lower than all the PPAC ones. Among these last ones, the results obtained in the first measurement with the tilted configuration are systematically higher than the other two datasets, with the difference becoming more pronounced above $\sim$500 MeV. It should be noticed, however, that all observed differences are well within the combined systematic uncertainties affecting the results here reported, as discussed below.

Each of the four datasets collected at n\_TOF represents a new result by itself, and should be considered independently from each other, for example in re-evaluating the fission cross section ratio. Nevertheless, for the purpose of this paper it may be convenient to combine all four datasets in a single one. To this end, a weighted average was performed, taking into account only the statistical errors (since the systematic errors are similar for the different datasets, in first approximation they can be neglected in the weighted average). The ratio obtained with this procedure can be more easily compared with data from previous measurements, as well as with evaluations and results of theoretical calculations. Furthermore, it can be used to analyze the dispersion of the n\_TOF datasets and draw further conclusions on their accuracy.

\begin{figure}[h!]
\includegraphics[angle=0,width=1.1\linewidth,keepaspectratio]{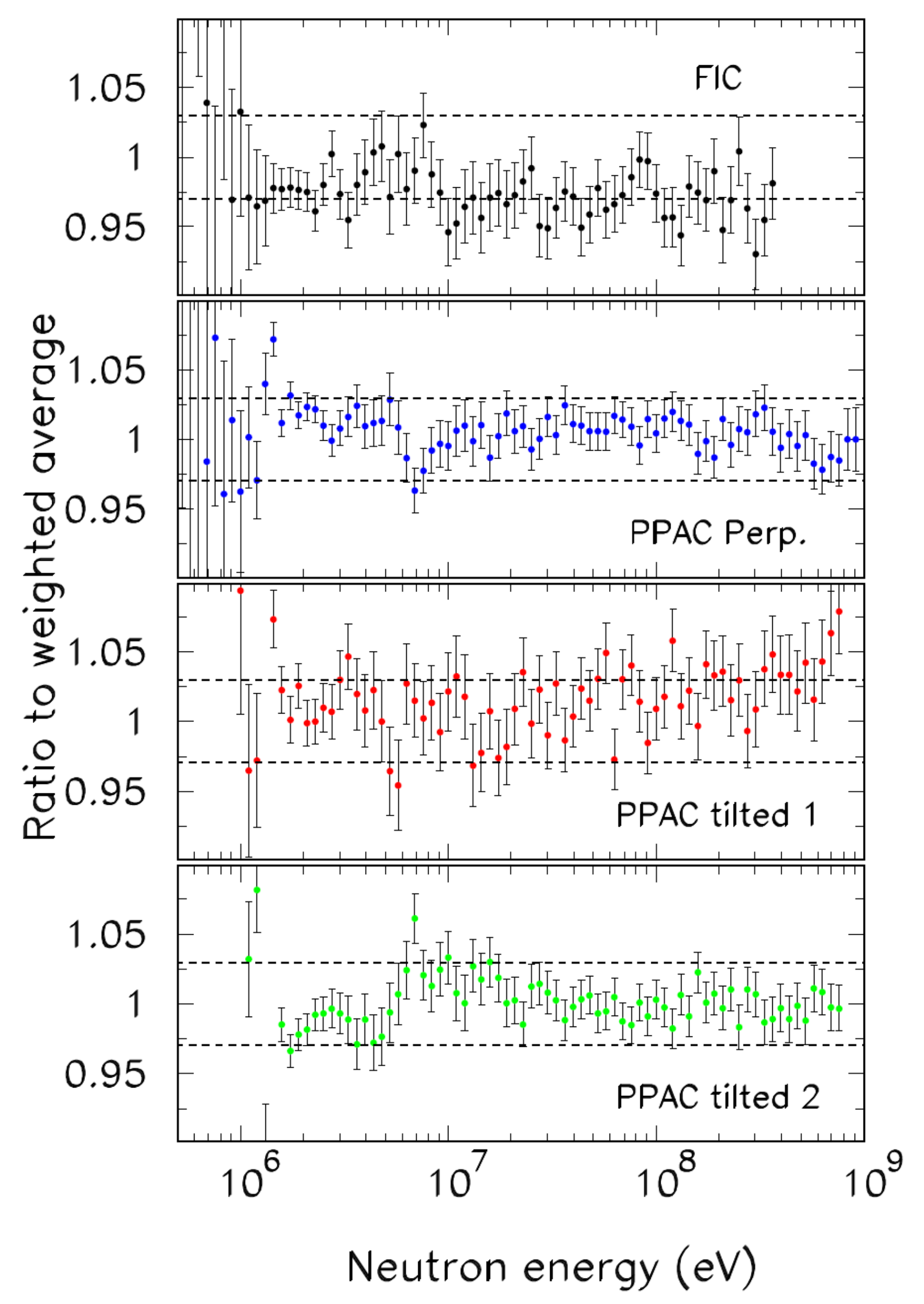}
\caption{\label{fig:ratioratio} (Color online) Ratio between the various datasets collected in the four different measurements at n\_TOF and their weighted average. The dashed lines represent the limit of the 3\% uncertainties.}
\end{figure}

This last point is illustrated in Figure~\ref{fig:ratioratio}, which shows the divergence of each n\_TOF result from their weighted average, as a function of the neutron energy. Except for some details, the difference in all cases is within the $\sim$3\% systematic uncertainty associated with each dataset (in the crude approximation adopted here, not taking into account the existing correlations between some datasets, the systematic uncertainty on the weighted average is around 1\% up to 200 MeV, and between 1 and 2\% above this energy). This observation provides further strength on the accuracy of the results here reported.

\begin{figure}[h!]
\includegraphics[angle=0,width=1.25\linewidth,keepaspectratio]{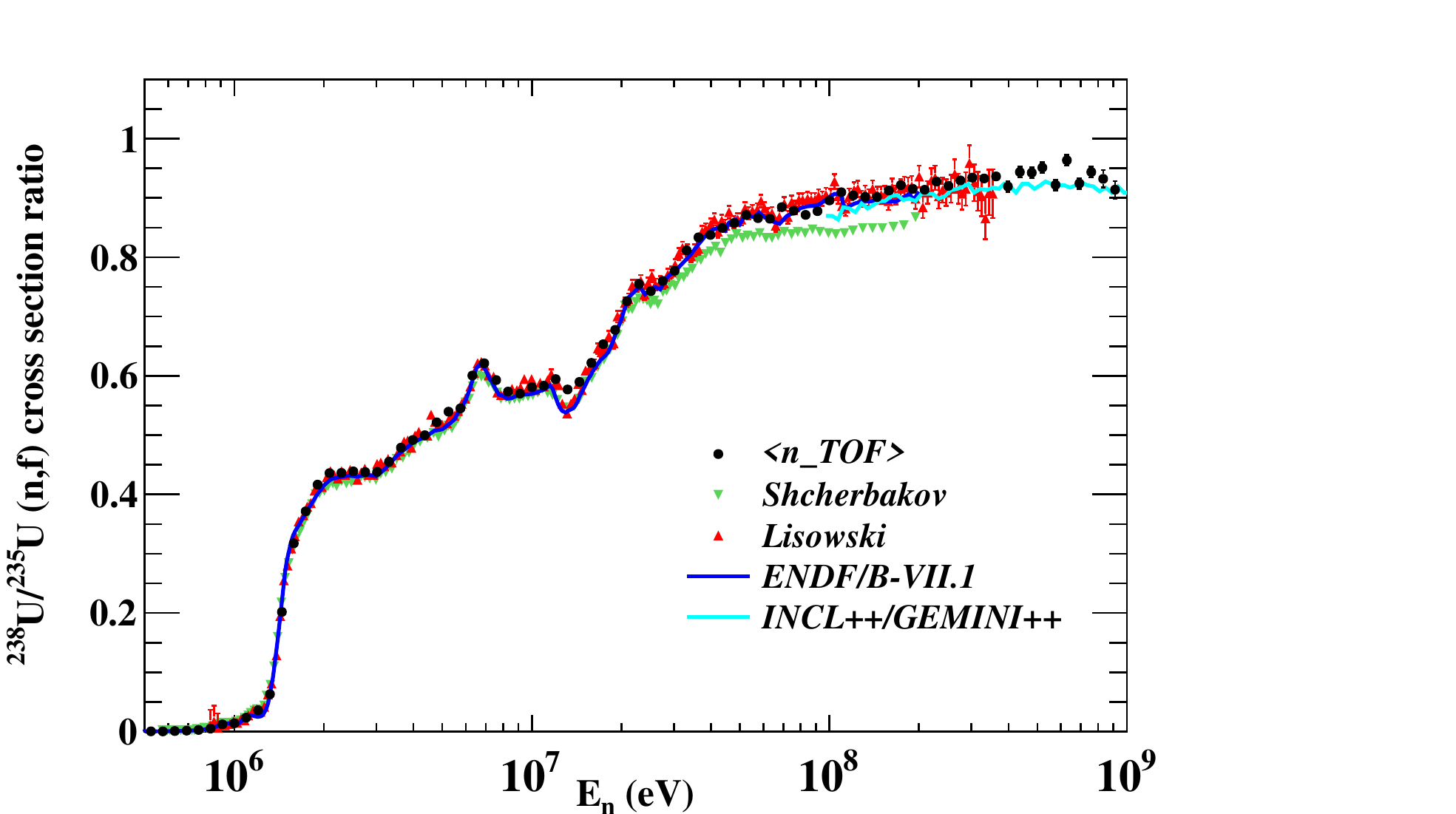}
\caption{\label{fig:rationtof} (Color online) $^{238}$U/$^{235}$U (\textit{n,f}) cross section ratio obtained from the weighted average of the four n\_TOF datasets compared with previous results from Ref.~\cite{liso} and~\cite{shcherba}, with the ratio extracted from the ENDF/B-VII.1 library and with the results of model calculations from Ref. \cite{Lo14}.}
\end{figure}

The unique n\_TOF fission cross section ratio discussed above is compared with the two major datasets extending to high energy in Figure~\ref{fig:rationtof} (the data from Nolte \textit{et al.}~\cite{nolte} are not reported in the figure due to their large errors). A very good agreement is observed with both the data from Lisowski \textit{et al.}~\cite{liso} and current ENDF/B-VII.1 evaluation all the way up to 200 MeV, definitely ruling out the lower values reported by Shcherbakov \textit{et al.}~\cite{shcherba}. Above this value, the n\_TOF data seem to maintain a constant or slightly increasing trend, contrary to the data of Lisowski \textit{et al.}, which seem to indicate the start of a decline in the higher energy points.

Some important information on the fission mechanism at high energy can be extracted from the comparison of the present data with the results of recent theoretical calculations of the fission cross sections of various actinides at high energy. More details on the calculations can be found in~\cite{Lo14}. The aim of the work was to provide a mean for estimating the fission cross section in the hundreds of MeV neutron energy range, in particular above 100 MeV, where no (or very few) reliable data exist. This is also the case for the fission cross section ratio discussed here. In the upper energy range from 200 MeV to 1 GeV, considered in this work, model calculations were performed in order to estimate the absolute cross sections and, consequently, their ratio, to be compared with the present data. In Ref. \cite{Lo14}, (\textit{p,f}) and  (\textit{n,f}) cross sections were evaluated for both isotopes in the energy range from 100 MeV to 1 GeV using the intranuclear cascade code INCL++~\cite{Bo13} coupled to the evaporation-fission code GEMINI++~\cite{Ma10}. Two parameters in the fission model were optimized so to reproduce the (\textit{p,f}) cross sections measured in Ref.~\cite{Ko06} from 200 MeV to 1 GeV (see Ref.~\cite{Lo14} for details). The corresponding (\textit{n,f}) cross sections were evaluated without further adjustment of model parameters, because the models already account for the observed differences in (p,f) and (n,f) cross sections around 200 MeV, while protons and neutrons are expected to have similar behavior with increasing incident energy and the fission cross sections to be of the same order of magnitude around 1 GeV. Such an assumption made it possible to reproduce also the (\textit{n,f}) data from 100 MeV to 200 MeV, even if the conditions of validity of the intranuclear cascade model are poorly satisfied at such relatively low energies. The theoretical cross sections are affected by statistical errors inherent in the Monte Carlo technique, which propagate to their ratios and have to be considered in the comparison with the experiment. The results of the model calculations are show in Figure~\ref{fig:rationtof} by the light-blue curve. A good agreement is observed at all energies. Although some systematic effects in the calculations may be at least partially compensated in the cross section ratio, the good agreement observed in the figure is an evidence that the theoretical treatment of Lo Meo \textit{et al.} is adequate for predicting fission cross sections of actinides above 100 MeV neutron energy.

\section{Conclusions}

The $^{238}$U/$^{235}$U fission cross section ratio has been measured at n\_TOF, for the first time up to 1 GeV, with two different detection systems, one of which used in two different geometrical configurations. The results of the different measurements agree with each other within the estimated systematic uncertainty on each dataset of approximately 3\%. Furthermore, the results are in all cases in agreement with evaluated cross sections and standard values between 0.5 and 20 MeV, thus providing confidence on their accuracy.
The different results have been combined in order to obtain a unique value of the ratio, which has then been compared with previous results and evaluations, up to 200 MeV, and with theoretical calculations based on intranuclear cascade model coupled with an evaporation-fission code, all the way up to 1 GeV. The extracted ratio rules out the results of Shcherbakov $\textit{et al.}$, and is in very good agreement both with current evaluation and standards, up to 200 MeV, and with theoretical calculations up to 1 GeV. These results could be used to improve the accuracy of current libraries, and in particular of the standard used in a variety of applications.

\begin{acknowledgments}
The research leading to these results has received funding from the European Atomic Energy Community’s (Euratom) Seventh Framework Program FP7/2007-2011 under the Project CHANDA (GA n. 605203). 
\end{acknowledgments}

\bibliography{u8u5}

\end{document}